\documentstyle[epsfig]{aa}

\newcommand{\be}{\begin{equation}}
\newcommand{\ee}{\end{equation}}

\begin{document}

\title{X-ray flashes from off-axis nonuniform jets}
\author{Z.P. Jin\inst{1,2} and D.M. Wei\inst{1,2}}
\institute{Purple Mountain Observatory, Chinese Academy of
Sciences, Nanjing, 210008, China \and National Astronomical
Observatories, Chinese Academy of Sciences, Beijing, 100012,
China}
\date{Received date/ Accepted date}

\abstract{It is widely believed that outflows of gamma-ray bursts
are jetted. Some people also suggest that the jets may have
structures like $\epsilon(\theta)\propto\theta^{-k}$. We test the
possibility of X-ray flashes coming from such jets in this paper.
Both qualitative and quantitative analyses have shown that this
model can reproduce most observational features for both X-ray
flashes and gamma-ray bursts. Using common parameters of gamma-ray
bursts, we have done both uniform and nonuniform jets' numerical
calculations for their fluxes, spectra and peak energies. It seems
that nonuniform jets are more appropriated to these observational
properties than uniform ones. We also give a spectrum and flux fit
to the most significant X-ray flash, XRF971019 by our model. We
also have shown that in our model the observational ratio of
gamma-ray bursts to X-ray flashes is about several.
} \maketitle

\section{Introduction}

X-ray flash(XRF) is a kind of recently identified explosion. Its
most properties are qualitatively similar to those of gamma-ray
burst(GRB) such as duration, temporal structure, spectrum and
spectral evolution except peak energy and flux. X-ray flash's peak
energy and flux are lower, but their distributions just smoothly
join the gamma-ray burst, there seems to be no obvious borderline
between XRF and GRB. These similarities led to the suggestion that
the X-ray flash is in fact "X-ray rich" gamma-ray burst (Kippen et
al. 2003), maybe they have same origins except for different
conditions.

The similarity between XRF and GRB suggests that the X-ray flash
might come from an off-axis nonuniform gamma-ray burst's
jet(Woosley et al. 2003; Rossi et al. 2002; Zhang \& Meszaros
2002b). When a burst is observed at the center of the jet, it will
be detected as a normal gamma-ray burst. But the burst tend to be
"dirty" when it is observed at a large viewing angle (Zhang \&
Meszaros 2002b), off-axis ejected matter takes less energy and has
lower Lorentz factor. So its $E_{\rm p}$ will shift to X-ray
responsibility, and it will be observed as an X-ray flash.

In this paper, we adopt a structured jet model where all the
energy and mass of ejected matter per unit solid angle and the
initial bulk Lorentz factor depend on the angle distance $\theta$
from the center as power laws
$\epsilon(\theta)\propto(\theta/\theta_{\rm c})^{-k}$, $m_{\rm
ej}(\theta)\propto(\theta/\theta_{\rm c})^{-k_{2}}$,
$\gamma(\theta)\propto(\theta/\theta_{\rm c})^{-k_{1}}$(Meszaros
et al. 1998; Rossi et al. 2002). We take $k=2$ for a nonuniform
jet, Rossi et al. have shown that $1.5\leq k\leq 2.2$ is the
reasonable value for fitting observations well(Rossi et al. 2002).

Frail et al.(2001) had given the jet angles distribution with
known redshift gamma-ray bursts. The jet's opening angle are range
from 0.05 to 0.4 rad(Frail et al. 2001), and most common value is
0.12 rad(Perna et al. 2003). The gamma-ray energies released are
narrowly clustered around $E_{\gamma}\sim 5\times10^{50}{\rm
ergs}$(Frail et al. 2001).

We introduce our model and give out some analytical solutions in
Sect.2. In Sect.3 we present numerical results of spectra and
fluxes for both uniform and nonuniform jets, and calculate the
gamma-ray bursts to X-ray flashes observational ratio. Finally, we
give a discussion and draw some conclusions in Sect.4.

\section{The Model}

We consider a relativistic outflow where the energy per unit solid
angle depend as power law on the angular distance from the center
$\theta$ (Meszaros et al. 1998; Zhang \& Meszaros 2002a; Rossi et
al. 2002): \be
\epsilon(\theta)=\left\{\begin{array}{ll}\epsilon_{\rm c} &
0\leq \theta \leq  \theta_{\rm c}\\
\epsilon_{\rm c}(\frac{\theta}{\theta_{\rm c}})^{-k} & \theta_{\rm
c}\leq
\theta \leq  \theta_{\rm j}\\
0 & \theta_{\rm j}\leq \theta\\
\end{array} \right . \ee
and the ejected matter per unit solid angle and the bulk Lorentz
factor also depend on $\theta$ as power laws: $m_{\rm
ej}(\theta)=m_{\rm ej}(0)(\theta/\theta_{\rm c})^{-k_{2}}$,
$\gamma(\theta)=\gamma(0)(\theta/\theta_{\rm c})^{-k_{1}}$
($\theta_{\rm c}\leq\theta\leq\theta_{\rm j}$). The deceleration
radius at $\theta$ is $r_{\rm
d}(\theta)=(\frac{3\epsilon(\theta)}{4\pi n\gamma(\theta)^{2}{\rm
m}_{\rm p}{\rm c}^{2}})^{1/3}=r_{\rm
d}(0)(\frac{\theta}{\theta_{\rm c}})^{(-k+2k_{1})/3}$ .

All our calculations will be done at the time when an outflow just
reaches its deceleration radius where the blast wave is formed.
Because of the beaming effect of large Lorentz factor at this
time, there is no obvious observation difference between isotropic
and anisotropic outflows. That means a jetted outflow with a
viewing angle $\theta_{\rm v}$ is observationally similar to an
isotropic outflow with bulk Lorentz factor
$\gamma=\gamma(\theta_{\rm v})$. So we can use the solutions from
an isotropic explosion model (Sari \& Piran 1999) to do an
analysis by choosing different Lorentz factor at different viewing
angle $\theta_{\rm v}$: \be \nu_{m}(\theta)=44{\rm
KeV}(\frac{\epsilon_{\rm e}}{0.1})^{2}(\frac{\epsilon_{\rm
B}}{0.1})^{1/2}(\frac{\gamma(\theta)}{300})^{4}n_{1}^{1/2} \ee
 \be F_{\nu,{\rm max}}(\theta)=220\mu{\rm Jy}D_{28}^{-2}(\frac{\epsilon_{\rm B}}{0.1})^{1/2}(\frac{\gamma(\theta)}{300})^{2}(\frac{r_{\rm d}(\theta)}{5.4\times10^{15}})^{3}n_{1}^{3/2}
\ee $(\nu F_{\nu})_{\rm max}(\theta)=\nu_{\rm max}(\theta)F_{\nu
,\rm max}(\theta)$ \be =11.2{\rm KeV}\cdot{\rm cm}^{-2}\cdot{\rm
s}^{-1}D_{28}^{-2}(\frac{\epsilon_{\rm
e}}{0.1})^{2}\frac{\epsilon_{\rm
B}}{0.1}(\frac{\gamma(\theta)}{300})^{6}(\frac{r_{\rm
d}(\theta)}{5.4\times10^{15}})^{3}n_{1}^{2} \ee

These equations describe the emission features from a shock
between outflows and external mediums. Generally a external shock
is not ideal for reproducing a highly variable burst(Sari et
al,1998), but it can reproduce a burst with several
peaks(Panaitescu \& Meszaros 1998) and may therefore be applicable
to the class of long, smooth bursts(Meszaros 1999).

Electrons in the external mediums will be accelerated to a power
law distribution. These electrons will product a broken power law
spectrum with photon spectrum indexes $\alpha$(low) and
$\beta$(high) in the range $-3/2$ to $-2/3$ and $-(p+1)/2$ to
$-(p+2)/2$ through synchrotron emission (Katz 1994; Cohen et al.
1997; Lloyd \& Petrosian 2000). Here $p$ is the power law index of
accelerated electrons. There is no difference for an isotropic
burst viewing from different direction, but for an anisotropic
burst, $E_{\rm p}$ changes from several KeV (or several eV,
depends on $\gamma(0)$, $\theta_{\rm j}$, $\theta_{\rm c}$ and
$k$) to hundreds KeV (or several MeV) at different viewing angle,
covers both GRBs and XRFs.

From Eq.(2) and Eq.(4), we get: \be \nu_{\rm
max}(\theta)\propto\gamma^{4}(\theta)\propto\gamma^{4}(0)(\frac{\theta}{\theta_{\rm
c}})^{-4k_{1}}\ee \be (\nu F_{\nu})_{\rm
max}(\theta)\propto\gamma^{6}(0)r_{\rm
d}^{3}(0)(\frac{\theta}{\theta_{\rm c}})^{-k-4k_{1}} \ee Compared
Eq.(5) and Eq.(6) we get \be F(\theta)\sim (\nu F_{\nu})_{\rm
max}(\theta)\propto \nu_{\rm max}^{\delta}(\theta) \ee Here,
$\delta=\frac{k+4k_{1}}{4k_{1}}$, only depends on the relation
between $k$ and $k_{1}$. When $k_{1}=k_{2}=k/2$ and $\gamma(0)$ is
a constant for every explosion, $\delta=3/2$, that is the simplest
solution. It will lead to the conclusion $E_{\rm p}\propto
L^{2/3}$, which is close to the observational relation $E_{\rm
p}\propto L^{1/2}$(Lloyd et al. 2000; Amati et al. 2002; Wei \&
Gao 2003).

For an isotropic jet when $\theta_{\rm v}>\theta_{\rm j}$,
$F_{\nu,{\rm max}}\propto(\gamma(1-\beta\cos(\theta_{\rm
v}-\theta_{\rm j})))^{-3}$, $\nu_{\rm
max}\propto(\gamma(1-\beta\cos(\theta_{\rm v}-\theta_{\rm
j})))^{-1}$(Yamazaki et al. 2002; Granot et al. 2002), in this
case we find $\delta$ is about 4 and $E_{\rm p}\propto L^{1/4}$.

Outflows with a lower Lorentz factor, which is called as a "dirty"
fireball or a failed gamma-ray burst, may also produce a X-ray
flash(Heise et al. 2003, Huang et al. 2002). It has the same
spectrum and flux as we have just given out for a nonuniform jet.
A dirty fireball will draw the same conclusion. It means that we
cannot distinguish our model from a dirty fireball model just by a
single X-ray flash. If simply assume that the bulk Lorentz factor
$\Gamma\propto L^{1/6}$, we get $E_{\rm p}\propto\Gamma^{4}\propto
L^{2/3}$. That means maybe we cannot distinguish nonuniform jet
model from dirty fireball model even by statistical properties.

Here we have to point out that Eq.(3) do not take cooling of
electrons into account which may cause the $F_{\nu,{\rm max}}$ to
decrease 1 to 3 magnitude. We will take the cooling effect into
account in our numerical calculation in the next section.

\begin{figure}
\epsfig{file=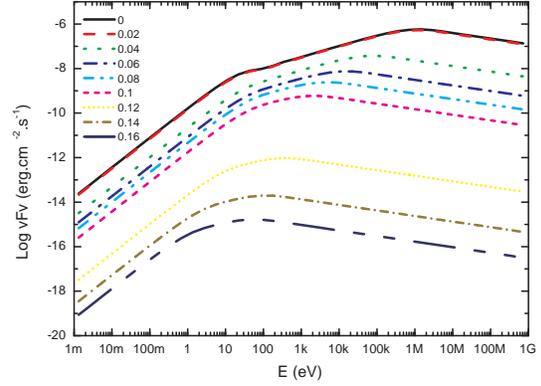,width=80mm} \caption{The spectra of a
non-uniform jet viewed from different viewing angles, here we
adopt $\theta_{\rm j}=0.1$, $\theta_{\rm c}=0.02$, $k=2$,
$k_{1}=k_{2}=k/2$, $r_{\rm d}=4.0\times 10^{16}{\rm cm}$,
$\gamma(0)=500$, $p=2.5$, and the viewing angles $\theta_{\rm v}$
change from 0 to 0.16.}
\end{figure}

\begin{figure}
\epsfig{file=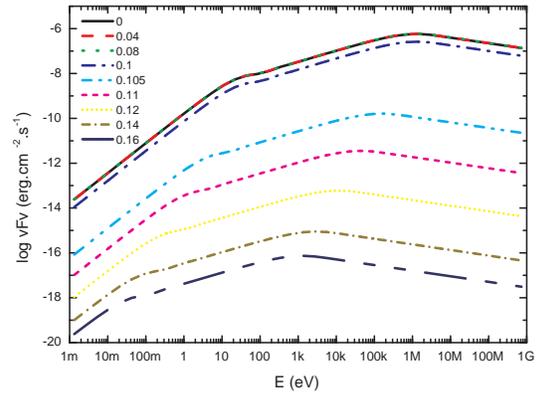, width=80mm} \caption{The spectra of an
uniform jet viewed from different viewing angles, here we adopt
$\theta_{\rm j}=0.1$, $r_{\rm d}=4.0\times 10^{16}{\rm cm}$,
$\gamma=500$, $p=2.5$, and the viewing angles $\theta_{\rm v}$
also change from 0 to 0.16. Here lines represent 0 to 0.08 are
overlapped.}
\end{figure}

\section{Numerical result}

We have given out some useful conclusions using a simplified
analysis. But for more realistic, the observed flux at frequency
$\nu$ is the integral of equal arrival time surface of a jet:

\be F_{\nu}=\int_{\gamma_{\rm min}}^{\gamma_{\rm max}}d\gamma_{\rm
e}\int_{0}^{2\pi}d\phi\int_{0}^{\theta_{\rm j}}\sin\theta d\theta
D ^{3} \frac{P'(\nu D^{-1})N(\gamma_{\rm e})}{4\pi d_{\rm
L}^{2}}\ee Here $P'(\nu ')=\frac{\sqrt{3}{\rm e}^{3}B}{{\rm
m}_{{\rm e}}{\rm c}}f(\chi)$ is synchrotron radiation power at
frequency $\nu '$ from a single electron in the fireball co-moving
frame (Rybicki \& Lightman 1979).
$D=(\gamma(\theta)(1-\beta\cos\Theta))^{-1}$ is the Doppler factor
translating from fireball co-moving frame to observer frame.
$\Theta$ is the angle between direction of outflow and
line-of-sight. The electrons' distributions can be written as(Dai
et al. 1999):

1.For $\gamma_{\rm c}\leq\gamma_{\rm e,mim}$ \be N(\gamma_{\rm
e})=\left\{\begin{array}{ll}C_{1}\gamma_{\rm e}^{-2} &
\gamma_{\rm c}\leq\gamma_{\rm e}\leq\gamma_{\rm e,min}\\
C_{2}\gamma_{\rm e}^{-(p+1)} & \gamma_{\rm e,min}\leq\gamma_{\rm
e}\leq\gamma_{\rm e,max}\\\end{array} \right . \ee

where

$C_{1}=C_{2}\gamma_{\rm min}^{-p+1}$,\\

$C_{2}=[\frac{\gamma_{\rm e,min}^{1-p}-\gamma_{\rm c}^{1-p}}{\gamma_{\rm c}(p-1)}+\frac{\gamma_{\rm c}^{-p}-\gamma_{\rm e,max}^{-p}}{p}]^{-1}N_{\rm e}$ \\

Where $N_{\rm e}=\frac{1}{3}r^{3}n_{1}$ is the total number of
electrons per unit solid angle, equals to the number of protons in
swept ISM. $\gamma_{\rm c}$ is the critical electron Lorentz
factor above which synchrotron radiation is significant(Sari et
al. 1998). \be \gamma_{\rm c}=\frac{3{\rm m}_{\rm
e}}{16\epsilon_{\rm B}\sigma_{\rm T}{\rm m}_{\rm p}{\rm
c}}\times\frac{1}{t\gamma^{3}n}=\frac{3{\rm m}_{\rm
e}}{8\sigma_{\rm T}{\rm m}_{\rm p}}\epsilon_{\rm
B}^{-1}\gamma(\theta)^{-1}r(\theta)^{-1}n_{1}^{-1} \ee

2.For $\gamma_{\rm e,min}\leq\gamma_{\rm c}\leq\gamma_{\rm e,max}$
\be N(\gamma_{\rm e})=\left\{\begin{array}{ll}C_{3}\gamma_{\rm
e}^{-p} &
\gamma_{\rm e,min}\leq\gamma_{\rm e}\leq\gamma_{\rm c}\\
C_{4}\gamma_{\rm e}^{-(p+1)} & \gamma_{\rm c}\leq\gamma_{\rm
e}\leq\gamma_{\rm e,max}\\\end{array} \right . \ee

where

$C_{3}=C_{4}\gamma_{\rm min}^{-p+1}$,

$C_{4}=[\gamma_{\rm c}^{-1}\gamma_{\rm
e,min}^{1-p}+\frac{(1-p)\gamma_{\rm e,min}^{-p}-\gamma_{\rm
e,max}^{-p}}{p}]^{-1}N_{\rm e}$

3.For $\gamma_{\rm c}\geq\gamma_{\rm e,max}$ \be
\begin{array}{ll} N(\gamma_{\rm e})=C_{5}\gamma_{\rm e}^{-p} &
\gamma_{\rm e,min}\leq\gamma_{\rm e}\leq\gamma_{\rm
e,max}\\\end{array}\ee

where

$C_{5}=\frac{p-1}{\gamma_{\rm e,min}^{1-p}-\gamma_{\rm
e,max}^{1-p}}N_{\rm e}$

We assume the bulk Lorentz factor $\gamma(\theta)$ keeps a
constant before the outflows arrive at their deceleration radius
$r_{\rm d}$. Here we choose $k_{1}=k_{2}$ which makes $r_{\rm d}$
a constant at different $\theta$, this assumption will make
calculations very simple. But notice that for different $\theta$,
the time for outflows arrive at $r_{\rm d}$ is different. We
calculate the flux when outflow reaches $r_{\rm d}$ at viewing
angle $\theta_{\rm v}$, that time is:

\be T=\frac{(1-\beta\cos({\rm max}[\theta_{\rm v}-\theta_{\rm
j},0]))r_{\rm d}}{\beta c} \ee

The equal arrival time surface at $\theta$ is:\be
r(\theta)=\frac{\beta cT}{1-\beta\cos\Theta} \ee

Our numerical results have been shown in Fig.1 and Fig.2. We
choose $\theta_{\rm j}=0.1$, $\theta_{\rm c}=0.02$,
$\gamma_{0}=500$, $r_{\rm d}=4.0\times 10^{16}{\rm cm}$, $d_{\rm
L}=1\times 10^{28}{\rm cm}$, $\epsilon_{\rm e}=0.1$,
$\epsilon_{\rm B}=0.1$, $p=2.5$, $k=2$ for a nonuniform jet and
$k=0$ for an uniform one.

It seems that for a nonuniform jet, the spectra and fluxes fit
both GRBs and XRFs observations fairly well. Viewed from the
center, $E_{\rm p}$ is about $100{\rm KeV}\sim 1{\rm MeV}$, and
flux is about $10^{-7\sim-6}{\rm erg}\cdot{\rm cm}^{-2}\cdot{\rm
s}^{-1}$, these are the typical values for GRBs(e.g. in Fig.1, the
cases for $\theta_{\rm v}=0,0.02,0.04$). While when viewed from
off-axis, $E_{\rm p}$ is about $10{\rm KeV}-100{\rm KeV}$, flux is
about $10^{-8\sim-7}{\rm erg}\cdot{\rm cm}^{-2}{\rm s}^{-1}$(e.g.
in Fig.1, the cases for $\theta_{\rm v}=0.04,0.06$), these are the
typical values for XRFs. When $\theta_{\rm v}=0.08,0.1$, $E_{\rm
p}$ is about a few KeV, the flux seems a little lower, but still
can be detected if the source distance is not so large.

But it seems that for an uniform jet, the flux from the jet edge,
where XRFs are thought to be from in this model, are too low. In
this case the spectra peak at 10-100KeV, the fluxes are about
$10^{-13\sim -10}{\rm erg}\cdot{\rm cm}^{-2}{\rm s}^{-1}$(e.g. in
Fig.2, the cases for $\theta_{\rm v}=0.105,0.11,0.12$). It can
only explain nearby XRFs, such as $z\leq 0.2$(Yamazaki et al.
2002). But XRFs are more likely have cosmological origins(Heise
2002).

\begin{figure}
\epsfig{file=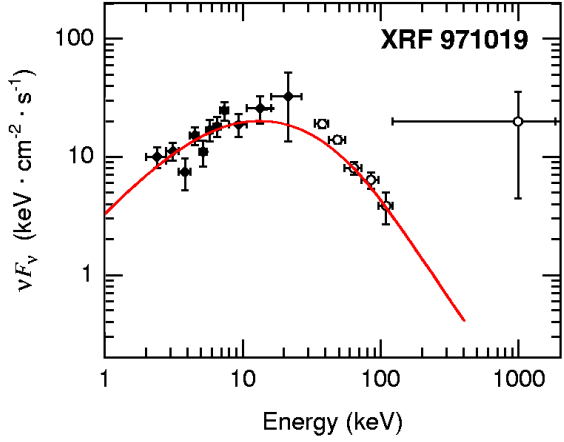, width=80mm} \caption{A spectrum and flux
fit to XRF971019 using our calculation program. Here we chose
$\theta_{\rm j}=0.1$, $\theta_{\rm c}=0.02$, $\gamma_{0}=500$,
$\theta_{\rm v}=0.08$, $p=5.4$, $r_{\rm d}=4.0\times 10^{15}{\rm
cm}$, $d_{\rm L}=4.2\times 10^{26}{\rm cm}\sim 136{\rm Mpc}$.}
\end{figure}

Fig.3 is a fit to the most significant X-ray flash XRF971019 using
our nonuniform jet model. Here we choose $p=5.4$ because the high
$\nu F_{\nu}$ power law index is about -1.7. This value seems a
little unusual, but similar to the case of GRB980425 whose
high-energy power-law photo index is $3.8\pm0.7$(Galama et al.
1998). XRF is defined as X-ray transients with duration less than
$1000{\rm s}$ which is detected by WFC(on BeppoSAX) but not
triggered GRBM(on BeppoSAX). Actually, this definition may lead to
strong selection effects on those transients with high power law
index.

We assume that every explosion has same luminosity and jet shape,
luminosity at center is $L_{\nu}(0)$, and at viewing angle
$\theta$ is $L_{\nu}(\theta)=L_{\nu}(0)(\theta/\theta_{\rm
c})^{-2k_{1}}$. Depending on the viewing angle $\theta$, a burst
can be detected only at the distance $D(\theta)$: \be
D(\theta)\leq D_{\rm
max}(\theta)=(\frac{L_{\nu}(0)(\theta/\theta_{\rm
c})^{-2k_{1}}}{4\pi F_{\nu,{\rm min}}})^{1/2} \ee Here
$F_{\nu,{\rm min}}$ is the threshold of a detector. So the numbers
of GRBs($N_{\rm GRBs}$) and XRFs($N_{\rm XRFs}$) are:
\begin{eqnarray*}N_{\rm GRBs}=\int_{0}^{\theta_{\rm c}}\frac{4}{3}\pi
D(\theta_{\rm c})^{3}n\frac{\sin(\theta)
d\theta}{2}\\+\int_{\theta_{\rm c}}^{\theta_{\rm
cr}}\frac{4}{3}\pi D(\theta)^{3}n\frac{\sin(\theta) d\theta}{2}
\end{eqnarray*}
\begin{eqnarray*} N_{\rm XRFs}=\int_{\theta_{\rm cr}}^{\theta_{\rm j}}\frac{4}{3}\pi
D(\theta)^{3}n\frac{\sin(\theta) d\theta}{2} \end{eqnarray*} $n$
is the number of bursts per unit volume, $\theta_{\rm cr}$ is the
critical angle, if viewing angle is larger than it the explosion
will be observed as XRFs. We assume the peak energy from the jet
axis $E_{\rm p,c}=1{\rm MeV}$, here we divide XRFs from GRBs with
$E_{\rm p}=90{\rm KeV}$, then the critical angle $\theta_{\rm
cr}=(\frac{E_{\rm p,c}}{90\rm KeV})^{\frac{1}{4k_{1}}}\theta_{\rm
c}$. We get: \be \frac{N_{\rm GRBs}}{N_{\rm
XRFs}}=\frac{\frac{3}{2}\theta_{\rm c}^{-1}-\theta_{\rm
cr}^{-1}}{\theta_{\rm cr}^{-1}-\theta_{\rm j}^{-1}}\sim 2.6 \ee

This result depends on the parameters we have chosen. Changing
these parameters in a reasonable range will change this result a
little but not much. On the other hand, these parameters can be
restricted by observed ratio.

\section{Discussion and conclusion}

It is more likely a gamma-ray burst outflow is jetted. The energy
distributions are more likely smoothly deducing with angle
distance from the center than uniform in the cone. In this paper,
we use a structured jet model which the energy per unit solid
angle decreasing as $\epsilon(\theta)\propto\theta^{-k}$. We
reproduce the key observational features of XRFs and GRBs. Our
calculations are based on the external shock model which seems not
ideal for reproducing GRBs with highly variable temporal
structure, but still adapt to those bursts with smooth light
curves. And our calculations should also adapt to the internal
shock model.

We have chosen $\theta_{\rm c}=0.02$ and $\gamma(0)=500$ in our
numerical calculations. But the lower limit to $\theta_{\rm c}$ is
about $1/\gamma_{\rm max}\sim 10^{-3}$, and the bulk Lorentz
factor in the center have an maximum value to $\gamma_{\rm
max}\sim 10^{5}$(Piran 1999; Rossi et al. 2002). Thus $E_{\rm p}$
may be in excess of MeV when the axis just point towards us though
the probability for this case is small.

We take $k=2$ for a nonuniform jet, Rossi et al. had shown that
$1.5\leq k\leq 2.2$ is the reasonable value for fitting
observations well(Rossi et al. 2002; Zhang \& Meszaros 2002a). For
the cases of $1\leq k\leq 3$, the calculated spectra and fluxes,
also the relation between $E_{\rm p}$ and $L$ are similar to that
of $k=2$. Because of beaming effect and shape of equal arrival
time surface, the radiation mainly comes from outflows just
pointing towards observer. We will observe a similar spectrum and
flux at a larger(smaller) viewing angle when $k$ is
smaller(larger). But it will change our solutions of GRBs to XRFs
observational ratio. We also assume $k_{1}=k_{2}=1$ in the
calculations through this paper, $k_{1}\neq k_{2}$ is a more
complex case which may lead to very complicated calculation.

We find that the observational ratio of GRBs to XRFs is about
several in our model. It predicts that more XRFs or soft GRBs will
be found in the future when more sensitive instruments are
launched. Barraud et al. have presented 35 GRBs/XRFs spectra from
HETE-2 whose band is 4-700KeV(Barraud et al. 2003), the numbers of
bursts with $E_{\rm p}$ higher and lower than 90KeV in these 35
GRBs/XRFs are 24 and 11, similar to our calculated result 2.6.

The observation had shown a correlation $E_{\rm p}\propto
L^{1/2}$(Lloyd et al. 2000; Amati et al. 2002; Wei \& Gao 2003).
This result is still not well explained except for some arguments
based on some very simple assumption(Lloyd et al. 2000). We find
in our model, for a single explosion viewed from different angles,
$E_{\rm p}\propto L^{2/3}$, close to the observational relation.
For a dirty fireball the relation between initial Lorentz factor
and explosion energy is uncertain. If simply assuming that
$\Gamma\propto L^{1/6}$, it leads to the same conclusion we have
presented.

Our model cannot be distinguished from a dirty fireball model just
by a single X-ray flash, but for these two models the statistical
properties such as the observational ratio of GRBs to XRFs are
different. In addition, the structured jet model is also different
from a dirty fireball in their afterglows. A very obvious feature
for a structured jet is that when $\theta_{\rm v}/\theta_{\rm c}$
and $k$ is large, there will be a prominent flattening in the
afterglow light curve, and a very sharp break occurring at the
time $\gamma\sim(\theta_{\rm v}+\theta_{\rm c})^{-1}$ after the
flattening(Rossi et al. 2002; Wei \& Jin 2003). We suggest that
such features are more likely to be found in an X-ray flash
afterglow rather than in a gamma-ray burst afterglow if an XRF
does come from an off-axis nonuniform jet, because an XRF has a
larger $\theta_{\rm v}/\theta_{\rm c}$ value than a GRB in this
model.

Orphan afterglows once caused great expectations to testify GRB
collimation(Rhoads 1997). In a nonuniform jet, orphan afterglows may be
generated in two ways. One way is when viewing angle is out of the jet
edge(e.g. in Fig.1 the cases $\theta_{\rm v}=0.12,0.14,0.16$, the
fluxes will increase to detectable values at later time). In this case,
the GRBs/XRFs are undetectable due to beaming effect but afterglows are
detectable which are less beamed. The other probable way is that the
Lorentz factor along the line of sight is sufficiently small that the
peak energy $E_{\rm p}$ is below the X-ray band. This case did not
appear in our calculations, it will appear if we choose a smaller value
of $\theta_{\rm c}/\theta_{\rm j}$ or a larger value of $k_{1}$.

We have neglected the evolution of bulk Lorentz factor and lateral
expansion of the jet which would make the calculations more
complex and we think these effects are not very important before
outflows arrive at their deceleration radius. Huang et al. have
given out an overall evolution of jetted gamma-ray bursts in
detail(Huang et al. 2000), we suggest that one should take all
these effects into account for more realistic calculations.

\acknowledgements This work is supported by the National Natural
Science Foundation (grants 10073022, 10225314 and 10233010) and
the National 973 Project on Fundamental Researches of China
(NKBRSF G19990754).

\end{document}